\documentclass[11pt]{article}

\usepackage[usenames,dvipsnames,table]{xcolor}

\usepackage{jheppub} 

\usepackage{graphicx}
\usepackage{amsmath, amssymb}
\usepackage{youngtab}
\usepackage{manfnt}
\usepackage{float}

\newcommand{\Sfour}{Spin-${\bf Z}_4$\ }

\newcommand{\be}{\begin{eqnarray}}
\newcommand{\ee}{\end{eqnarray}}
\newcommand{\bea}{\begin{eqnarray}}
\newcommand{\eea}{\end{eqnarray}}
\def\eqn{\eqref}

\newcommand{\nn}{\nonumber}
\newcommand{\bn}{\begin{enumerate}}
\newcommand{\en}{\end{enumerate}}

\title{Gapped Chiral Fermions}

\author[a]{Shlomo S. Razamat}
\author[b]{and David Tong}

\affiliation[a]{Department of Physics, Technion, Haifa, 32000, Israel}
\affiliation[b]{Department of Applied Mathematics and Theoretical Physics \\ University Cambridge, CB3 0WA, UK}

\emailAdd{razamat@physics.technion.ac.il}
\emailAdd{d.tong@damtp.cam.ac.uk}

\abstract{In principle, there is no obstacle to gapping fermions preserving any global symmetry that does not suffer a 't Hooft anomaly. In practice, preserving a symmetry that is realised on fermions in a chiral manner necessitates some dynamics beyond simple quadratic mass terms. We show how this can be achieved using familiar results about the strong coupling dynamics of supersymmetric gauge theories and, in particular, the phenomenon of confinement without chiral symmetry breaking. We present simple models  that gap  fermions while preserving a symmetry group under which they transform in chiral representations. For example, we show how to gap a collection of 4d fermions that carry the quantum numbers of one generation of the Standard Model, but without breaking electroweak symmetry. We further  show how to gap fermions in groups of 16 while preserving certain discrete symmetries that exhibit a  mod 16 anomaly.}

\begin{document} 

\maketitle
\flushbottom

\section{Introduction}

What symmetries are lost when fermions gain a mass? Naively, one might think that chiral symmetries are broken, while vector-like symmetries survive. This is certainly the case if one simply writes down a quadratic mass term for fermions. Moreover, it is often the case if the fermions get their mass through some strong coupling effect -- say, a four-fermion term in $d=1+1$ dimensions, or a confining gauge theory in $d=3+1$ -- where the chiral symmetry is typically  broken spontaneously.

Typically, but not always. The purpose of this paper is to describe a number of simple models that give fermions a mass while preserving chiral symmetries. These will include both continuous symmetries and more subtle discrete symmetries. The phenomenon of gapping fermions while preserving a chiral symmetry sometimes goes by the name of {\it symmetric mass generation}. 

The real obstacle to giving fermions a mass while preserving a global symmetry $G$ is the 't Hooft anomaly associated to $G$ \cite{thooftanomaly}. If the anomaly is non-vanishing, then the fermions cannot be trivially gapped without breaking $G$\footnote{Sometimes they can be ``non-trivially gapped", meaning they leave behind a topological quantum field theory that saturates the anomaly.}. However, if the anomaly vanishes then there is, in principle, no obstacle to gapping the  fermions while preserving $G$, even if they sit in a chiral representation.  The question is: how do we do it in practice?

It will be useful to have two simple examples in mind as we proceed, both of them in $d=3+1$ dimensions:

\begin{itemize}
\item Consider 15 Weyl fermions  carrying the quantum numbers of a single generation of the Standard Model under the symmetry $G=SU(3)\times SU(2) \times U(1)$. (We review these quantum numbers in Section \ref{smsec}.) Famously, the anomalies vanish. Can these fermions be gapped without breaking $G$?
\item Consider 16 Weyl fermions that enjoy a $G={\rm Spin-}{\bf Z}_4$ symmetry, meaning that the generator $U$ obeys $U^2 = (-1)^F$. Such a symmetry has a mod 16 anomaly and this  vanishes if $U$ acts on all 16 fermions in the same way, say by multiplying them by $+i$. The Spin-${\bf Z}_4$ symmetry prohibits a quadratic mass term but, with vanishing anomaly, there is nothing that prohibits the entire cohort of 16 to be gapped en masse. How can we achieve this?
\end{itemize}

For continuous symmetries, there is a long literature of proposals designed to gap chiral fermions  within the context of lattice gauge theory, starting with the insightful work of Eichten and Preskill \cite{eichten,creutz,poppitz,wen,request,demarco,kik,ww1,ww2,simon}. (A closely related discussion in the context of quantum Hall edge states can be found in \cite{haldane,levin}.) A common theme among these papers is that fermions can be  gapped while preserving chiral symmetries through the use of higher dimension operators.

These higher dimension operators are irrelevant. In a continuum field theory, if one starts with free fermions and adds only irrelevant interaction terms then obviously they will not gap the system. However, with an underlying lattice one can turn on irrelevant operators with a large coefficient so that the system is strongly coupled in the UV. In such a situation,  these irrelevant operators can dominate the physics, giving the fermions a mass comparable to the UV cut-off 

For someone steeped in the Wilsonian perspective on continuum quantum field theory, relying on dynamics at the UV cut-off to drive the low-energy physics of interest might induce a level of anxiety. Any such nervousness is likely to be compounded by the observation that, on closer inspection, the Eichten-Preskill mechanism seems not to work, with no hint of the gapped chiral phase appearing as one explores some (admittedly finite dimensional) parameter space  \cite{golt,poppitz2}.

In contrast, our interest in this paper lies firmly in the continuum. We do not allow ourselves to rely on strongly coupled UV physics. Instead, we wish to stay relevant. The purpose of the paper is to present a method to gap fermions, preserving a chiral symmetry, 
 by introducing new degrees of freedom, turning on relevant operators and flowing to a gapped phase in the infra-red.

In $d=3+1$ dimensions, the only relevant interactions involve non-Abelian gauge dynamics. As we will see, in many cases the gapped chiral phase can be achieved through a phenomenon that has long been understood: confinement without chiral symmetry breaking, sometimes referred to as {\it s-confinement}. This is a phenomenon that is best understood in supersymmetric theories where the first examples were given by Seiberg \cite{nati1,nati2}\footnote{The authors of \cite{tachyon1} previously advocated the use of supersymmetric gauge dynamics to explore symmetric mass generation. Their interest was in gapping 16 fermions  in $d=2+1$ dimensions while preserving time reversal, albeit viewed from the perspective of the bulk  $d=3+1$ dimensional SPT phase.}. Usually in s-confining theories one has  massless fermions in both the UV and IR, but with the 't Hooft anomalies for unbroken symmetries realised in startlingly different fashions. As we will explain, a small tweak of this idea allows us to gap fermions preserving chiral symmetries, including the example highlighted above of fermions in the Standard Model\footnote{A related  proposal to use gauge theories to drive symmetric mass generation was made in \cite{you1,you2}. The idea was that one could use gauge dynamics to flow to an interacting critical point with the hope that the multi-fermion operator, that was irrelevant in the ultra-violet, becomes relevant and can now be employed to gap the system. The theories we study here are similar in spirit, but  significantly simpler since the theory flows to a free critical point, which can subsequently be gapped.}.

\subsubsection*{The Plan of the Paper}

We start in Section \ref{4dsec} by presenting the basic idea, relating symmetric mass generation in $d=3+1$ dimensions to s-confinement. We then proceed to give a number of examples. In particular, in Section \ref{smsec}, we explain how to gap the fermions in a single generation of the Standard Model while preserving the chiral $SU(3) \times SU(2) \times U(1)$ symmetry.

In Sections  \ref{spin4sec} and \ref{3dsec}, we turn to discrete symmetries. In Section \ref{spin4sec} we describe how s-confinement also provides a mechanism to gap fermions in groups of 16, preserving a  \Sfour symmetry as described above. In Section \ref{3dsec}, we describe a novel supersymmetric theory in $d=2+1$ dimensions that is trivially gapped while preserving time reversal. We check that the corresponding mod 16 index is indeed vanishing, as it should be.

\section{Gapping Chiral Fermions in $d=3+1$}\label{4dsec}

In this section, we present a number of models that gap fermions in $d=3+1$ while preserving continuous symmetries that are realised in a chiral manner. The basic idea is very straightforward and, as explained in the introduction, follows from the phenomenon of confinement without chiral symmetry breaking, sometimes called s-confinement. 

\subsubsection*{The Rules of the Game}

First, let us spell out more clearly what we wish to achieve. We start with a collection of free, massless fermions, transforming in some anomaly free representation of a global symmetry group $G$. Our goal is to gap the fermions, preserving $G$.

Adding a quadratic mass term to the Lagrangian typically breaks $G$, while four-fermion terms are irrelevant.  This means that to achieve our goal we must add new degrees of freedom and allow them to interact with our original fermions while preserving $G$. These new degrees of freedom can include scalars, fermions and gauge bosons. But, with each, come a number of caveats. 

First, the scalars. These can transform  in any representation of $G$ since they can be trivially gapped and decoupled from the system without breaking $G$. However, if the scalars do transform under $G$ then we must take care to ensure that they don't condense, spontaneously breaking $G$. 

In contrast, if the fermions transform under $G$, then it must be in a vector-like representation. (Obviously it would be cheating if we simply added fermions in the conjugate representation of $G$, gap the whole system and declare victory.) Insisting that any additional fermions transform in a vector-like representation ensures that they can be trivially decoupled by giving them a quadratic mass term, preserving $G$.

Finally, if we wish to drive some strong coupling dynamics in the infra-red (and we do) then we must also add gauge bosons\footnote{In principle, it may be possible to induce symmetric mass generation in lower dimensions without gauge interactions. In the Appendix, we show that this is not possible in supersymmetric Wess-Zumino models with four supercharges (i.e. the dimensional reductions of $4d$ ${\cal N}=1$).}.  Crucially, we are not allowed to gauge the global symmetry $G$ that we care about: this is to remain a global symmetry of the interacting theory.\footnote{The symmetry $G$ will remain unbroken and non-anomalous throughout,  which means that there is nothing to stop us repeating the discussion with $G$ now a low-energy gauge symmetry. Nonetheless, for pedagogical purposes it is simplest to first think of $G$ as a global symmetry. We'll discuss this further in Section \ref{smsec} where we discuss gapping fermions that carry Standard Model quantum numbers  without breaking $G=SU(3) \times SU(2) \times U(1)$.}
%
%
%
 However, if the enlarged system of scalars and fermions enjoys a second symmetry, $H$, then we may consider gauging it. We require both that $G$ commutes with $H$ and, moreover, that $G$ and $H$ have no mixed 't Hooft anomaly.

Our final requirement is that there exists a regime of parameter space where the gauge bosons decouple. This can be achieved by including scalars that can fully Higgs the gauge group $H$. This means that one can take a limit where we are left only with the original massless fermions of interest, with all other degrees of freedom heavy. 

The upshot of these rules is that our original chiral theory interacts with an auxiliary vector-like theory such that it is straightforward to decouple the vector-like matter, leaving behind the original massless fermions. However, instead we will  tune parameters so that we  bring down the heavy, vector-like degrees of freedom until they interact with the light fermions, gapping the entire system, all while leaving $G$ untouched. That is the goal. As we now explain, the properties of theories exhibiting s-confinement provide exactly what we need. 

\subsubsection*{From S-Confinement to Symmetric Mass Generation}

In any confining theory, the  fundamental quarks are bound together in the infra-red to form mesons, baryons and other composites. We will be interested in confining theories that enjoy a global symmetry $G$. In confining theories -- and in contrast to our preceding discussion --  it is often the case that $G$ has a  't Hooft anomaly. If $G$ is to survive the RG flow to the infra-red unscathed,  then the spectrum of confined particles must include massless states that replicate the 't Hooft anomaly. If this is not possible, then $G$ must be spontaneously broken.

Long ago, 't Hooft argued that, in QCD with massless quarks, there is no spectrum of massless composite states that can replicate the anomaly  for the chiral symmetry \cite{thooftanomaly}. In other words, in QCD confinement implies chiral symmetry breaking.

However, there are other theories, closely related to QCD, where the 't Hooft anomalies can be matched by a massless composites in the infra-red. In this situation, it is possible that the theory exhibits confinement without breaking the chiral symmetry $G$. Although it is possible to find putative examples of this phenomenon without invoking supersymmetry, the addition of supersymmetry  provides the extra control required to be confident of the low-energy physics. In the supersymmetric context, confinement without chiral symmetry breaking is referred to as s-confinement, with ``s" for ``smooth".

In s-confining theories, the action of $G$ is realised differently on the fundamental fermions $Q$ in the UV, and the composite states in the IR which, for now, we refer to collectively as $M$. We then couple both UV and IR theories to a new sector, consisting of free fields $\widetilde{M}$ that transform in the representation of $G$ that is conjugate to  $M$.  This is achieved by turning on a superpotential term that, schematically, takes the form 
\be {\cal W}\sim \widetilde{M}M\label{flipoff}\ee
Couplings of this type have been previously considered in, for example, \cite{Barnes:2004jj,Dimofte:2011ju,Benvenuti:2017lle} and are sometimes referred to as ``flipping" the operator $M$.

From the perspective of the IR, we have not achieved anything surprising. The composite fermions $M$  can be viewed as fields in the IR and the superpotential  above is a mass term that gaps the system.
 However, from the perspective of the original gauge theory, we have quarks $Q$ and singlets $\widetilde{M}$ that typically transform in a chiral representation of $G$ but, by construction, one with vanishing 't Hooft anomaly. We can identify the coupling in the UV that replicates the  infra-red superpotential \eqn{flipoff}, and thus we have succeeded in gapping the fermions while preserving a global, chiral symmetry.

The coupling between the fundamental fermions $Q$ and singlets $\widetilde{M}$ will turn out to be irrelevant or marginally irrelevant in the ultra-violet. However, this is different from the situation described in the introduction where irrelevant operators are introduced on the lattice to gap chiral fermions. We do not need to turn on these irrelevant operators with a large coefficient, because they are examples of {\it dangerously irrelevant} operators: after the RG flow initiated by the gauge interactions they become relevant. Indeed, from the infra-red perspective, they are simply mass terms.

\subsection{A Non-Supersymmetric Warm Up}\label{su5sec}

We illustrate the general idea with a simple non-supersymmetric example. Consider the global symmetry 
\be G = U(1)\nn\ee
with a collection of 16 Weyl fermions with chiral charges under $G$ given by $3^{[5]},(-1)^{[10]}$ and $-5$, where the superscripts are multiplicities. This is an anomaly free representation.

In addition to the $G=U(1)$ symmetry, the 16 fermions enjoy an $H=SU(5)$ symmetry, under which those fermions with charge 3 transform in the $\bar{\bf 5}$ and those with charge $-1$ transform  in the ${\bf 10}$.  Importantly, $H$ and $G$ commute and have no mixed 't Hooft anomaly. 

We now gauge $H$. (Following the ``rules of the game" above, we should also introduce Higgs fields that can remove the gauge bosons, but when these scalars are heavy they do not affect the story.) We now have a familiar situation: an $SU(5)$ gauge theory, coupled to a fermion $\psi$ in $\bar{\bf 5}$, a fermion $\chi$ in the ${\bf 10}$, and a singlet fermion that we call $\widetilde{\zeta}$. 

 The infra-red dynamics of this chiral, non-Abelian gauge  theory is not known for sure but there is a good candidate, first proposed in \cite{raby}:  the gauge theory is thought to confine, with the $\psi$ and $\chi$ fermions combining into the massless, gauge invariant composite
\be \zeta \sim \psi\psi\chi\nn\ee
This has $G=U(1)$ charge $+5$. We assume that this indeed is the correct dynamics.

We now add a four-fermion interaction in the UV, 
\be {\cal L} \sim \widetilde{\zeta} \psi\psi\chi\nn\ee
This is a dangerously irrelevant operator. It  is irrelevant in the UV but, assuming the strong coupling dynamics described above, descends to a simple mass term $\widetilde{\zeta}\zeta$ in the IR, where it gaps the theory.

Hence, we have succeeded in gapping the fermions preserving the chiral $G=U(1)$ symmetry. The quantum numbers of the fermions under this $U(1)$ are rather artificial looking and, of course, were constructed by working backwards from the known dynamics of the $SU(5)$ gauge theory.  
In the rest of this section, we describe models which implement symmetric mass generation for simpler and more interesting chiral representations of global symmetries.

\subsection{$SU(N)$ with an Anti-Symmetric}\label{antisun}

Consider the global symmetry group
\be G = SU(N)\nn\ee
A chiral, anomaly free representation can be constructed from  a Weyl fermion $\widetilde{\chi}$ transforming in the anti-symmetric representation $\raisebox{-0.7ex}{\epsfxsize=0.12in\epsfbox{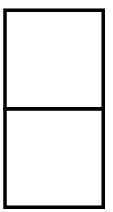}}$ and $N-4$ Weyl fermions $\psi$, each transforming in the anti-fundamental representation $\overline{\Box}$. The 't Hooft anomaly is well known to vanish, since the anomaly coefficients are
\be A(\overline\Box)  = -1\ \ \ {\rm and} \ \ \ A\left(\raisebox{-1.1ex}{\epsfxsize=0.12in\epsfbox{y7.eps}}\right) = N-4\nn\ee
Note that for $N=5$, this is closely related to the $SU(5)$ gauge theory described in Section \ref{su5sec}. However, the context is different: here we require that $SU(N)$ is a global symmetry, not a gauge symmetry. 

\subsubsection*{The Case of $N$ Even}

We start by discussing the case of $N$ even and $N\geq 6$. (The case of $SU(4)$ with an anti-symmetric is equivalent to $SO(6)$ with a ${\bf 6}$ and so is a vector-like theory in disguise.) We write  
\be N=2r\nn\ee
Our goal is to gap the fermions $\widetilde{\chi}$ and $\psi$, while preserving $G$.

First note that, in addition to the $G=SU(N)$ global symmetry, there is a further $H=SU(N-4) = SU(2r-4)$ symmetry which rotates the anti-fundamental fermions. To gap the  theory, we  gauge an $Sp(r-2)\subset H$ subgroup. We then add further scalars and fermions to endow the theory with ${\cal N}=1$ supersymmetry. The field and symmetry content of the theory is given by
\begin{table}[h!]
\begin{center}
\begin{tabular}{|c||c||c|c|}
  \hline 
   Field & $Sp(r-2)$ & $SU(N)$ & $U(1)_R$  \\
 \hline
 $Q$  & ${\bf 2(r-2)}$ & $\overline{\Box}$  & $2/N$  \\ 
 $\widetilde{M}$ & {\bf 1} & \raisebox{-0.7ex}{\epsfxsize=0.1in\epsfbox{y7.eps}} & $2-4/N$\\
\hline
 \end{tabular}
 \end{center}
\end{table}

\noindent
Here $U(1)_R$ is an  R-symmetry that arises from the introduction of the new, auxiliary scalar fields\footnote{For the readers who are not familiar with supersymmetry, the R-symmetry plays a special role in the study of these theories. By definition, the gaugino $\lambda$ has R-charge $+1$. The scalar fields transform with the R-charge specified in the table, while the associated fermions have $R[{\rm fermion}] = R[{\rm scalar}] -1$. The superpotential preserves the R-symmetry if it has charge $R[{\cal W}] = 2$.}.  We will discuss this further below.

Our original fermions $\psi$ and $\widetilde{\chi}$ inhabit the chiral superfields $Q$ and $\widetilde{M}$ respectively. They are now accompanied by  scalar superpartners, with the same transformation properties under both $Sp(r-2)$ and $SU(N)$. Furthermore, we have added a gaugino  in the adjoint of $Sp(r-2)$. Importantly, this gaugino is a singlet under $G=SU(N)$ and so, by the rules described previously, constitutes a legal addition to our theory. 

Finally, we add an interaction between our original fermions and the newly introduced scalars, in the guise of a superpotential
\be {\cal W}_{UV} = \widetilde{M}_{ij} Q^iQ^j\label{w1}\ee
where $i,j=1,\ldots, N$ are indices for the global symmetry $G$. In this expression $Q^iQ^j$ is contracted using the invariant symplectic form of $Sp(r-2)$ to yield a gauge invariant meson. 
This superpotential identifies the symmetries rotating $Q$ and $\widetilde M$ as appearing in the table above.  As anticipated, this superpotential is marginally irrelevant in the UV. This will no longer be the case as we flow to the IR.

As we explained previously, we wish the theory to have a regime in which the gauge group is fully Higgsed, so the gauge bosons and other extraneous fields become heavy, leaving us with only the massless fermions of interest. This can be achieved in theory above, but only at the expense of breaking supersymmetry. One first turns off the Yukawa terms arising from \eqn{w1}, decoupling the scalar quarks from their supersymmetric partners, so that the scalars are no longer obliged to transform under $G$.  We then give expectation values to $r-2$ of the scalars, ensuring that the gauge group is broken. The remaining scalars and the gaugino are then decoupled through mass terms, preserving the symmetry $G$. A similar process will work for all subsequent examples that we will  meet\footnote{There is a secondary question regarding this procedure: is it possible to accomplish the decoupling in a smooth fashion, or will the theory undergo a first order phase transition? Relatedly, if we softly break supersymmetry, do the results survive or is $G$ now spontaneously broken? We do not attempt to answer these questions here.}.

So far we have shown only that it is possible to return to the original free, massless fermions while preserving $G$. But our real goal is to understand how to gap these fermions preserving $G$. This follows automatically from the dynamics at the supersymmetric point, as studied by  Intriligator and Pouliot \cite{ip}. (The special case of $Sp(1) = SU(2)$ was previously considered by Seiberg \cite{nati1,nati2}.) The $Sp(r-2)$ gauge theory is an example of an s-confining theory, flowing in the infra-red, at the origin of the moduli space,  to a collection of massless mesons described by the composite field
\be M^{ij} = Q^iQ^j\nn\ee
This meson field transforms in the conjugate anti-symmetric representation $\overline{\raisebox{-0.7ex}{\epsfxsize=0.1in\epsfbox{y7.eps}}}$. The fact that the 't Hooft anomalies for $SU(N)$ and $U(1)_R$ match between the UV and IR provides compelling evidence for this result.

 The singlet $\widetilde{M}$ remains unaffected by the gauge dynamics, but the 
 ultra-violet superpotential \eqn{w1} descends to a more mundane mass term, 
 \be {\cal W}_{IR} = \widetilde{M}_{ij}M^{ij}\label{w2}\ee
 This is now a relevant operator, gapping the system while preserving the symmetry $G$. Indeed, as advertised above, from the perspective of the infra-red, the manner in which the fermions get a mass is neither chiral nor mysterious.  The magic happened in the strong coupling dynamics, and the fact that this theory exhibits confinement without chiral symmetry breaking. 
 
 Before we proceed, it is worth passing comment on the R-symmetry. The charges of the superfields are such that both $Sp(r-2)^2\cdot U(1)_R$ and $SU(N)^2\cdot U(1)_R$ anomalies cancel. (The former, of course,  is a requirement for the $U(1)_R$ symmetry to be a symmetry at all.) Since the theory is invariant under $U(1)_R$, one might wonder whether we have succeeded in demonstrating symmetric mass generation for $SU(5)\times U(1)_R$, rather than just $SU(5)$. There is a sense in which this is the case, but not for our original fermions $\psi$ and $\chi$. Indeed, the $R$ and $R^3$ 't Hooft anomalies are non-vanishing when restricted to $\psi$ and $\chi$, so they cannot be gapped preserving $U(1)_R$ without some help. In the present context, that help comes from the gaugino  $\lambda$, which has R-charge $+1$, and ensures that the full theory specified above has vanishing anomaly for both $R$ and $R^3$.

\subsubsection*{The case of $N$ Odd}

It is simple enough to generalise the above discussion to $N$ odd.  We again start with global symmetry $G=SU(N)$, with  a single Weyl fermion  $\widetilde{\chi}$ in the anti-symmetric representation $\raisebox{-0.7ex}{\epsfxsize=0.1in\epsfbox{y7.eps}}$ and $N-4$ Weyl fermions  $\psi$ in the anti-fundamental $\overline{\Box}$. This includes the case of $SU(5)$, coupled with a ${\bf 10}$ and $\bar{\bf 5}$, familiar from grand unification.

Since $N$ is odd, we now write
\be N = 2r-1\nn\ee
This time, we start by adding  extra fermions: first, we add  pair of fermions in conjugate representations of $G$: $\rho$ in the anti-fundamental $\overline{\Box}$ and $\widetilde{\rho}$ in the fundamental ${\Box}$. Taking $\psi$ and $\rho$ together, we have $N-3$ fermions in the $\overline{\Box}$ of $G$, and an $H=SU(N-3) = SU(2r-4)$ symmetry that rotates them. We are now in a similar situation to before and could try to gauge an $Sp(r-2) \subset H$ subgroup. 

Here we hit a snag; since $N$ is odd, we would be gauging $Sp(r-2)$ with an odd number of fundamentals and this suffers from the Witten anomaly. To avoid this, we add yet more fermions $\xi$, $2(r-2)$ of them, singlets under $G$ but transforming in the fundamental of $Sp(r-2)$. This cancels the Witten anomaly.  After  supersymmetrising the whole affair, the resulting field and symmetry content is given in the following table.

\begin{table}[h!]
\begin{center}
\begin{tabular}{|c||c||c|c|c|}
  \hline 
   Field & $Sp(r-2)$ & $SU(N)$ & $U(1)_R$  & $U(1)_A$\\
 \hline
 $Q$  & ${\bf 2(r-2)}$ & $\overline{\Box}$  & $2/(N+1)$  & 1\\ 
$S$ &  ${\bf 2(r-2)}$ & ${\bf 1}$  & $2/(N+1)$ & $-N$\\
 $\widetilde{P}$ & ${\bf 1}$ & $\Box$ & $2(N-1)/(N+1)$ & $N-1$ \\
 $\widetilde{M}$ & {\bf 1} & \raisebox{-0.7ex}{\epsfxsize=0.1in\epsfbox{y7.eps}} & $2(N-1)/(N+1)$ & $-2$\\
\hline
 \end{tabular}
 \end{center}
\end{table}

\noindent
Here the $Q$ multiplets now contain both the original $\psi$ fermions and the auxiliary fermion $\rho$. The $G$-singlet $\xi$ fermions are contained in $S$, while the $G$-fundamental $\widetilde{\rho}$ fermion is contained in $\widetilde{P}$. Finally, $\widetilde{M}$ contains our original fermion $\widetilde{\chi}$ as before.  We then add the superpotential
\be {\cal W}_{UV} = \widetilde{M}_{ij} Q^i Q^j + \widetilde{P}_i Q^i S\label{w3}\ee
with $i,j=1,\ldots,N$ the flavour indices for $G=SU(N)$. 
Once again, the gauge group flows to the infra-red and confines, resulting in gapless meson states without breaking the $G=SU(N)$ symmetry. These meson states are
\be M^{ij} = Q^iQ^j\ \ \ {\rm and}\ \ \ P^{i} = Q^i S\nn\ee
The $M^{ij}$ transform in the conjugate anti-symmetric representation $\overline{\raisebox{-0.7ex}{\epsfxsize=0.1in\epsfbox{y7.eps}}}$
of $G=SU(N)$, while $P^i$ transform in the anti-fundamental $\overline{\Box}$. By now the story should be familiar: the UV superpotential descends to the infra-red to
\be {\cal W}_{IR} = \widetilde{M}_{ij}M^{ij} + \widetilde{P}_iP^j\nn\ee
where it ensures that all states are gapped, preserving $G$.

The same remarks that we made about $U(1)_R$ in the case of $N$ even apply here too, both to $U(1)_R$ and the global symmetry $U(1)_A$. Both are free from 't Hooft anomalies, but only because of the contribution from the auxiliary fermions that we added along the way.

\subsubsection*{Breaking Supersymmetry}

The examples above rely on known results about supersymmetric gauge theories. We invoke supersymmetry only as a crutch to give us the requisite control over the strong coupling dynamics. It is natural to ask: can we achieve something similar without supersymmetry?

In general, our understanding of strongly coupled, non-supersymmetric gauge theories is not sufficiently advanced to give a definitive answer to this question. However, as we now explain, the symmetric gapped phase does survive soft breaking of supersymmetry.

To begin the discussion, consider  a supersymmetric theory that exhibits s-confinement. The low-energy degrees  consists of massless composite fermions -- whether mesons or baryons -- that saturate the 't Hooft anomaly and their scalar superpartners. Although the scalars transform under $G$, they do not condense at the origin of their moduli space and so $G$ is unbroken . However, this phase may be fragile. The concern is that any supersymmetry breaking deformation, no matter how small, may induce tachyonic masses  for the scalar mesons, rendering the origin of the moduli space unstable. After supersymmetry breaking, the ground state of the system would then  break $G$. A number of studies of softly broken supersymmetry in this context include \cite{susyb1,susyb2,susyb3}. 
 
However, despite first appearances, this does not immediately nullify the symmetric mass generation mechanism that we advocated above. The theory still confines and the ultra-violet Yukawa couplings  \eqn{w1}  still descend to infra-red mass terms of the form \eqn{w2}. Now there are two competing mass scales in the game. The first is the soft supersymmetry breaking scale $\mu$. We must take this to be $\mu \ll \Lambda$, where $\Lambda$ is the strong coupling scale of the supersymmetric gauge theory.  This hierarchy  ensures that we can still use  Seiberg duality as a good starting point for the infra-red physics. The second mass scale is $m$, the dynamically generated gap in the theory. This is of order $m\sim \Lambda$; the two differ only by a dimensionless Yukawa coupling.

In the absence of supersymmetry breaking, all fields have a gap $m$. Suppose that when we softly break supersymmetry by a UV scale $\mu$, the massless scalars pick up a tachyonic contribution $-\tilde{\mu}^2$ to their mass. Their full mass is then
\be m_{\rm scalar}^2 \sim m^2 - \tilde{\mu}^2\label{susybreaking}\ee
and this remains positive for suitably small $\tilde{\mu}$. Thus we see that symmetric mass generation persists for small supersymmetry breaking.

Of course, this had to be the case. The purpose of symmetric mass generation is to gap the system while preserving a symmetry $G$ which, in the current setting, is continuous. But such a phase is necessarily robust. An arbitrarily small perturbation cannot spontaneously break $G$ when the system is gapped since this would result in gapless Goldstone modes, in contradiction with the smooth variation of the spectrum. The symmetry breaking only occurs only if we perturb the system by an amount comparable to the gap. This is expectation is reflected in  \eqn{susybreaking}.

\subsection{The Standard Model}\label{smsec}

The Standard Model presents  a particularly interesting example of a non-anomalous chiral symmetry, with group 
\be  G = \frac{SU(3) \times SU(2) \times U(1)_Y}{{\bf Z}_6}\nn\ee
The anomaly free matter content consists of 15 right-handed Weyl fermions, sitting in representations of $G$ given by
\be   l^c_L: \  ({\bf 1},{\bf 2})_{-3} \ \  \, \  q^c_L: \  (\bar{\bf 3},{\bf 2})_{+1}\  \ ,\  \ e_R: \  ({\bf 1},{\bf 1})_{+6}  \  \ ,\   \ u_R:\ ({\bf 3},{\bf 1})_{-4}  \  \ ,\ \    d_R:\  ({\bf 3},{\bf 1})_{+2}  \nn\ee
We have rescaled the hypercharges to be integers. Note that we have not yet introduced the right-handed neutrino: it will make an appearance shortly.

In the previous examples, we viewed $G$ as a global symmetry\footnote{If $G$ is viewed as a global symmetry, rather than a gauge symmetry,  then the ${\bf Z}_6$ quotient is mandatory if the symmetry is to act faithfully on the fermion content. If it $G$ is a gauge symmetry, the ${\bf Z}_6$ quotient is optional. This issue is not important for the goal of symmetric mass generation but a broader discussion can be found in \cite{me}.}. Obviously, in the context of the Standard Model $G$ is a gauge symmetry and this means that it comes its own  dynamical scale,  $\Lambda_{\rm SM}$. Here,  $\Lambda_{\rm SM}$ could be viewed as either the the weak scale where the Higgs mechanism takes place, or the strong scale of confinement. 
The analog of symmetric mass generation is now finding a mechanism that gives  the fermions a mass $m$ that is independent of the scale $\Lambda_{SM}$. In particular, we should be able to give the fermions a mass $m\gg \Lambda_{\rm SM}$, where $G$ is weakly coupled, without spontaneously breaking  $G$.

To achieve this, we again introduce new degrees of freedom.  As in the previous example, the first step is to introduce yet further fermions that sit in vector-like representations of $G$. We write the original fermions in black (omitting their names),  with three additional pairs of fermions in red, 
\be   && ({\bf 1},{\bf 2})_{-3} \  \  \ \  (\bar{\bf 3},{\bf 2})_{+1}\ \ \ \ ({\bf 1},{\bf 1})_{+6}  \  \  \  \ ({\bf 3},{\bf 1})_{-4}\  \   \  \ ({\bf 3},{\bf 1})_{+2} \ \ \ \ {\color{red}({\bf 1},{\bf 1})_0} 
 \nn\\ 
&& {\color{red}({\bf 1},{\bf 2})_{-3} } \ \ \ \ \ \ \ \ \ \ \ \ \ \ \ \ \ \ \ \ \ \ \ \ \  \ \ \ \ \ \ \ \ \ \ \ \ \ \ \ \ \ \  \ \ \ \  {\color{red}({\bf 3},{\bf 1})_{+2} }\ \ \ \  {\color{red}({\bf 1},{\bf 1})_{0} }
\nn\\
&& {\color{red}({\bf 1},{\bf 2})_{+3}} \ \ \ \ \ \ \ \ \ \ \ \ \ \ \ \ \ \ \ \ \ \ \ \ \  \ \ \ \ \ \ \ \ \ \ \ \ \ \ \ \ \ \  \ \ \ \  {\color{red}(\bar{{\bf 3}},{\bf 1})_{-2} }
 \nn\ee
Crucially, the additional fermions sit in vector-like representations of $G$; it is trivial to give masses to each of the pairs without breaking $G$.  Note that we have added two fermions that are singlets under $G$; one of these can play the role of the right-handed neutrino.

The additional fermions mean that we have three pairs with the same quantum numbers: these are the fermions that sit in the first two lines above. The next step is to introduce an $H=SU(2)$ gauge symmetry (not to be confused with the $SU(2)$ global symmetry in $G$) under which these pairs of fermions transform as a doublet. Importantly, this symmetry does not have a mixed anomaly with $U(1)_Y$, so $G$ remains intact once we gauge $H$. The upshot is that we have a collection of fermions transforming as:
\begin{table}[H]
\begin{center}
\begin{tabular}{|c||c||c|c|c|}
  \hline 
   Fermion & $SU(2)_{\rm gauge}$ & $SU(3)$ & $SU(2)$  & $U(1)_Y$\\
 \hline
 $l$  & ${\bf 2}$ & ${\bf 1}$  & ${\bf 2}$  & $-3$\\ 
  ${l}'$  & ${\bf 1}$ & ${\bf 1}$  & ${\bf 2}$  & $+3$\\ 
$q$ &  ${\bf 1}$ & $\bar{\bf 3}$  & ${\bf 2}$ & $+1$\\
 $e$ & ${\bf 1}$ & ${\bf 1}$ & ${\bf 1}$ & $+6$ \\
 $u$ & ${\bf 1}$ & ${\bf 3}$ & ${\bf 1}$ & $-4$\\
 $d$ &  ${\bf 2}$ &  ${\bf 3}$ & ${\bf 1}$ & $+2$ \\
 ${d}'$ &  ${\bf 1}$ &  $\bar{\bf 3}$ & ${\bf 1}$ & $-2$ \\
 $\nu$  &  ${\bf 2}$ & ${\bf 1}$ & ${\bf 1}$ & 0\\
\hline
 \end{tabular}
 \end{center}
\end{table}

\noindent
At this stage, we introduce yet more fields to construct a supersymmetric extension of this model. These are scalar superpartners for each fermion listed above, together with a gaugino in the adjoint of $SU(2)_{\rm gauge}$. The end result is a collection of chiral multiplets, transforming as:

\begin{table}[H]
\begin{center}
\begin{tabular}{|c||c||c|c|c||c|c|}
  \hline 
   Field & $SU(2)_{\rm gauge}$ & $SU(3)$ & $SU(2)$  & $U(1)_Y$ & $U(1)_A$ & $U(1)_R$ \\
 \hline
 $L$  & ${\bf 2}$ & ${\bf 1}$  & ${\bf 2}$  & $-3$ & 0 & 0 \\ 
  ${L}'$  & ${\bf 1}$ & ${\bf 1}$  & ${\bf 2}$  & $+3$ & 3 & 2\\ 
$Q$ &  ${\bf 1}$ & $\bar{\bf 3}$  & ${\bf 2}$ & $+1$& $-1$ & 4/3\\
 $E$ & ${\bf 1}$ & ${\bf 1}$ & ${\bf 1}$ & $+6$ & 0 & 2\\
 $U$ & ${\bf 1}$ & ${\bf 3}$ & ${\bf 1}$ & $-4$ & $-2$ & 2/3\\
 $D$ &  ${\bf 2}$ &  ${\bf 3}$ & ${\bf 1}$ & $+2$ & $1$ & 2/3\\
 ${D}'$ &  ${\bf 1}$ &  $\bar{\bf 3}$ & ${\bf 1}$ & $-2$ & 2 & 4/3 \\
 $N$  &  ${\bf 2}$ & ${\bf 1}$ & ${\bf 1}$ & 0 & $-3$ & 0 \\
\hline
 \end{tabular}
 \end{center}
\end{table}

\noindent
where the additional fields from supersymmetry mean that the theory enjoys two further symmetries, $U(1)_A$ and $U(1)_R$.  One can check that the R-symmetry acts on the fermions  in  $L$, $Q$, $E$, $U$ and $D$ as the familiar $B-L$ symmetry of the Standard Model.

All the symmetries listed are preserved by  the gauge invariant superpotential
\be {\cal W}_{UV} = \epsilon_{ab}L^aL^bE + \epsilon_{ijk}D^iD^j U^k + \epsilon_{ab}L^aD^iQ_i^b  + \epsilon_{ab}L^aN L^{\prime b} + D^iN{D}'_i  \label{w4}\ee
where now $a,b=1,2$ are indices for $SU(2)\subset G$ and $i,j=1,2,3$ and indices for $SU(3)\subset G$. It is simple to check that each of these terms is invariant under $G$.

From hereon, the story is familiar. The strong coupling dynamics consists of an $SU(2)$ supersymmetric gauge theory coupled to six doublets: 2 in $L$, 3 in $D$ and $N$. This theory is known to exhibit s-confinement \cite{nati1,nati2} and, in the infra-red is described by a collection of 15 meson fields,
\be \widetilde{E} = \epsilon_{ab}L^a L^b \ \ , \ \widetilde{U}_k = \epsilon_{ijk}D^iD^j\ \ ,\ \ \widetilde{Q}^i_b = \epsilon_{ab}L^a D^i\ \ ,\ \ \widetilde{L}_b = \epsilon_{ab}L^a N \ \ , \ \ \widetilde{D}^i = D^iN \nn\ee
The superpotential \eqn{w4} descends to the infra-red where it becomes a collection of mass terms. 
\be  {\cal W}_{IR} = \widetilde{E}E + \widetilde{U}_kU^k + \widetilde{Q}_b^iQ_i^b  + \widetilde{L}^b L^{\prime b} + \widetilde{D}_i{D}'_i  \nn\ee
All  fields are gapped, preserving $G$.

\subsection{Further Generalisations}

Connoisseurs of supersymmetric gauge theories will have no trouble generalising these results to other chiral, anomaly free models using the many known s-confining theories \cite{son,Cho:1996bi,Csaki:1996eu,csaki,Spiridonov:2009za,Spiridonov:2011hf}. Here we briefly describe a few examples. 

At heart, the example of the Standard Model described above was constructed by embedding chiral representations of $SU(3)\times SU(2)\times U(1)_Y$ into 
\be G= SU(6)\ \mbox{with}\ \raisebox{-1.1ex}{\epsfxsize=0.12in\epsfbox{y7.eps}}\ \mbox{and 2 $\overline{\Box}$}\nn\ee
through the more familiar grand unified embedding into $SU(5) \subset SU(6)$.  Symmetric mass generation was then realised by viewing $G$ as the global symmetry of an $SU(2)$ gauge theory with six fundamental chirals and its (conjugate) singlet mesons. A slightly more complicated route realises  $G$ through an $Sp(n)$ gauge theory, with  six fundamentals and a traceless anti-symmetric, again accompanied by its mesons. This theory is known to s-confine and, for $n\geq 2$, preserves an $G=SU(6) \times U(1)$ symmetry \cite{Cho:1996bi,Csaki:1996eu}.

Another interesting, anomaly free chiral representation is given by
\be G = SU(N)\ \mbox{with}\ \overline{\raisebox{-1.1ex}{\epsfxsize=0.12in\epsfbox{y7.eps}}} \ {\rm and}\ 
\raisebox{-0.2ex}{\epsfxsize=0.22in\epsfbox{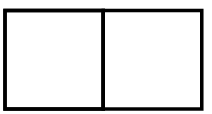}}\ \mbox{and 8 $\overline{\Box}$}\nn\ee
In addition to $G$, the fermions have an $H=SU(8)$ symmetry that acts on the anti-fundamentals. For $N=5$, we may gauge a $G_2\subset SO(7) \subset H$ symmetry. that acts on 7 of the 8 anti-fundamentals After suitable supersymmetrisation, the theory s-confines, yields a meson spectrum consisting of a $\raisebox{-1.1ex}{\epsfxsize=0.12in\epsfbox{y7.eps}}$, a $\overline{\raisebox{-0.2ex}{\epsfxsize=0.22in\epsfbox{y2.eps}}}$, and a $\Box$, which can then be paired with the gauge singlet fermions to gap the system \cite{csaki}.

Relatedly, for $N=6$ we may gauge a ${\rm Spin}(7) \subset H$ symmetry, with the 8 anti-fundamentals transforming in the spinor representation.  This results in a meson spectrum consisting of $\raisebox{-1.1ex}{\epsfxsize=0.12in\epsfbox{y7.eps}}$ and $\overline{\raisebox{-0.2ex}{\epsfxsize=0.22in\epsfbox{y2.eps}}}$, which again can be paired with the gauge singlets \cite{csaki}.

\section{A Spin-${\bf Z}_4$ Symmetry and the Mod 16 Anomaly}\label{spin4sec}

In recent years, there has been impressive progress in understanding 't Hooft anomalies associated to discrete symmetries. These anomalies are associated to cobordism groups \cite{df,kap} and underlie the classification of symmetry protected topological, or SPT, phases. Often, these discrete  anomalies are valued in ${\bf Z}_N$ for some $N$. This means that  fermions can be gapped, preserving the symmetry, only in groups of $N$. 

For example, in $d=0+1$ and $d=1+1$, Majorana fermions can be gapped in groups of 8 while preserving a suitable discrete symmetry, as first shown in the pioneering work of Fidkowski and Kitaev \cite{fk,ryu,qi}. (This discrete symmetry is time reversal with $T^2=+1$ in $d=0+1$, and chiral fermion parity $(-1)^{F_L}$ in $d=1+1$; for a review of the triality symmetry that underlies these calculations, see \cite{review}.)

In $d=3+1$ dimensions, the analogous question is how to gap fermions preserving a Spin-${\bf Z}_4$ symmetry. Such a  symmetry has a generator $U$ which obeys
\be U^2 = (-1)^F\nn\ee
This means that any scalar must transform as $\pm 1$, while any  Weyl fermion must transform as $\pm i$ under the ${\bf Z}_4$. 

There is a remarkable mod 16 anomaly associated to such a Spin-${\bf Z}_4$ symmetry. We first perform suitable conjugations so that all Weyl fermions are right-handed. Then the anomaly is given by
\be \nu_4 = n_+ - n_-\ \ \ \ {\rm mod}\ 16\label{nu4}\ee
where $n_\pm$ count the number of fermions that transform as $\pm i$. The fact that 16 Weyl fermions are special was first noted in \cite{bentov}; the concrete statement about the Spin-${\bf Z}_4$ symmetry and its relationship to the cobordism group $\Omega_5^{{\rm Spin}-{\bf Z}_4} = {\bf Z}_{16}$ was  stated in \cite{tachyon2,garcia}. 

The Spin-${\bf Z}_4$ symmetry prohibits quadratic mass terms for fermions. The question is: can we find a non-perturbative mechanism that lifts Weyl fermions in groups of 16? This would be the four-dimensional analog of the Fidkowski-Kitaev mechanism for lifting Majorana fermions in low dimensions in groups of 8.

In fact, as we now show, several of the examples from the previous section have this property. In these cases, the \Sfour symmetry is embedded in a continuous group, so does not provide new information beyond the perturbative anomalies. (The interplay between perturbative and non-perturbative anomalies was studied  in \cite{elitzur,nakarin}.) However, in many cases one can break these continuous symmetries -- say, by adding irrelevant 4-fermion terms to the action --  leaving behind only the \Sfour of interest. Indeed, the Standard Model  itself has a \Sfour symmetry, which acts as a combination of hypercharge and B-L \cite{garcia}. (Further discussions of the \Sfour symmetry in the context of the Standard Model can be found in \cite{hsieh,nak,juv1,juv2}.)  If one augments the Standard Model with all possible higher dimension operators (see, for example, \cite{higher} for a list of dimension six operators) then B-L is broken, but the \Sfour symmetry remains. 

Viewed this way, the non-supersymmetric $SU(5)$ chiral gauge theory described in Section \ref{su5sec} provides a particularly simple example where there is a Spin-${\bf Z}_4 \subset U(1)$ symmetry. In that case, it trivially multiplies all fermions by $i$. The UV theory has 16 fermions, and so $\nu_4=0$ as it must since, as we have seen, the theory is gapped while preserving \Sfour. 

\subsubsection*{Supersymmetry and the \Sfour R-symmetry}

Because the \Sfour symmetry acts differently on bosons and fermions, when embedded in a supersymmetric theory it must be a \Sfour R-symmetry.

Usually we normalise the R-symmetry so that the gaugino has charge $R[\lambda] = +1$, while chiral multiplets typically have fractional charge. For our purposes, it is better to multiply by the common denominator so that all charges are integer. We can then embed a \Sfour symmetry inside $U(1)_R$ if the gaugino has odd charge, while all chiral multiplets have even charge. Recall that the fermions in the chiral multiplet have $R[{\rm fermion}] = R[{\rm scalar}] -1$, so this ensures that all bosons have even charge while fermions have odd charge. Performing a $U(1)_R$ rotation by $e^{i\pi R/2}$ will then act as a \Sfour symmetry.

In what follows, we take the gaugino to transform as
\be \mbox{\Sfour}: \lambda \rightarrow i \lambda\nn\ee
The transformation of the fermions in a chiral multiplet $Q$ depends on whether the scalar is even or odd. If we denote the fermion in $Q$ as $\psi$, then we have
\be \mbox{\Sfour}: Q \rightarrow  \pm Q\ \ \ \Rightarrow \ \ \psi\rightarrow \mp i\psi \nn\ee
This ensures that the gaugino Yukawa couplings are invariant. To preserve Spin-${\bf Z}_4$, the superpotential must be odd. (This can be viewed as  cancelling the minus sign that comes from the $d^2\theta$ measure over superspace).

\subsubsection*{Examples: $SU(N)$ with an Anti-Symmetric}

A glance at the tables of $U(1)_R$ charges in Section \ref{4dsec}  will reveal that none of them have a \Sfour subgroup. However, it is not difficult to find such subgroups embedded within both $U(1)_R$ and the global symmetries.

Let's return to our simplest example from Section \ref{antisun} with global symmetry $G=SU(N)$, a Weyl fermion in ${\raisebox{-0.7ex}{\epsfxsize=0.1in\epsfbox{y7.eps}}}$, and $N-4$ Weyl fermions in $\overline{\Box}$. As we saw, the analysis is slightly different for $N$ odd and $N$ even. We will find that the embedding of the \Sfour R-symmetry is different in these two cases.

The story is simplest for $N$ odd. Here it is straightforward to embed
\be \mbox{\Sfour} \subset U(1)_R \times U(1)_A\nn\ee
To achieve this, we simply need to rotate in  $U(1)_R$ by $\pi/2$, and in $U(1)_A$ by $-\pi/(N+1)$. The resulting transformation of the various chiral multiplets is given by

\begin{table}[H]
\begin{center}
\begin{tabular}{c|cccc}
    &  $Q$ & $S$ & $\widetilde{P}$ & $\widetilde{M}$ 
    \\  \hline
\Sfour &  $+1$ &$-1$ & $+1$  & $-1$ 
 \end{tabular}
 \end{center}
\end{table}
\noindent
We don't need to count the index mod 16 since a straightforward calculation shows that there are equal numbers of fermions transforming as $\pm i$ so we have, simply, 
\be \nu_4=0\nn\ee
Things are more interesting when  $N$ is even. This time we  wish to find an embedding of 
\be \mbox{\Sfour} \subset SU(N) \times U(1)_R\nn\ee
To do this, we can  augment a  $U(1)_R$ rotation of $\pi/2$ by the following $SU(N)$ transformation
\be {\rm diag}(\underbrace{\omega,\ldots, \omega}_{N/2},\underbrace{\omega^{N+1},\ldots,\omega^{N+1}}_{N/2})\ \ \ {\rm with}\ \ \omega^{2N} =1\nn\ee
This  transformation has unit determinant, and hence sits inside $SU(N)$,  only when
\be N=2r\ \ {\rm with}\ \ r\ {\rm odd}\nn\ee
In this case, the $N$ chiral multiplets $Q$ split into two sets, each of $N/2$, which we denote as $Q$ and $Q'$. Similarly, the mesons split into three sets, $\widetilde{M}$ and $\widetilde{M}''$ each of dimension $\frac{1}{8} N (N-2)$ and $\widetilde{M}'$ of dimension $\frac{1}{4}N^2$. The theory is invariant under a \Sfour symmetry with
\begin{table}[H]
\begin{center}
\begin{tabular}{c|ccccc}
    &  $Q$ & $Q'$ & $\widetilde{M}$ & $\widetilde{M}''$ & $\widetilde{M}'$
    \\  \hline
\Sfour &  $+1$ &$-1$ & $-1$  & $-1$ & $+1$ 
 \end{tabular}
 \end{center}
\end{table}

\noindent
 If one tries to implement these transformations in the theory with $N=2r$ when $r$ is even, it turns out that they are embedded in an anomalous $U(1)$ symmetry, and so are not  symmetries of the theory. (The study of anomalous discrete symmetries, following from their embedding in continuous groups, was initiated in \cite{iross}.)

We can now calculate the mod 16 anomaly. Clearly the $Q$ and $Q'$ cancel in their contribution. The remaining fields yield
\be \nu_4 &=& \frac{1}{2}(N-4)(N-3) + \frac{1}{8}N(N-2) + \frac{1}{8} N(N-2) - \frac{1}{4} N^2  \nn\\ &=& \frac{1}{2}(N^2 - 8 N + 12)\nn\ee
where, in the first line,  the terms arise from the gaugino (using ${\rm dim}(Sp(n)) = n(2n+1)$) and $\widetilde{M}$, $\widetilde{M}''$ and $\widetilde{M}'$ respectively.  It is simple to check that 
\be \nu_4 = 0\ \ {\rm mod} \ 16\ \ \mbox{whenever $N=2r$ with $r$ odd}\nn\ee
We learn that the mod 16 anomaly vanishes, as indeed it must for any trivially gapped theory.

\section{Time Reversal in $d=2+1$ and the Mod 16 Anomaly}\label{3dsec}

In $d=2+1$ dimensions, there is no meaning to left- and right-handed fermions. Nonetheless, a more subtle notion of chirality exists depending on how fermions transform under time reversal. For a Majorana fermion $\chi$, there are two options which differ by a sign
\be {\rm T}: \chi \rightarrow \pm \gamma^0 \chi\label{tr}\ee
This obeys
\be {\rm T}^2 = (-1)^F\nn\ee
Theories with such a time-reversal exhibit a mod 16 anomaly given by \cite{fcv,senthil,witten,parity}
\be {\nu}_3 = \widetilde{n}_+ - \widetilde{n}_- \ \ {\rm mod \ 16}\nn\ee
where $\widetilde{n}_\pm$ counts the number of Majorana fermions that transform with a $\pm$ sign under time reversal \eqn{tr}. A non-vanishing ${\nu}_3$ can be viewed as an obstruction to placing the theory on an unoriented manifold with ${\rm Pin}^+$ structure \cite{kap}.

If the theory also has a $U(1)$ symmetry, then it is more convenient to work with Dirac fermions.  In a basis in which all gamma matrices are real, these can be written as  $\psi = \chi_1 +i\chi_2$, with $\chi_1$ and $\chi_2$ Majorana fermions. If we choose $T$ to act identically on each Majorana, say as $\chi_i\rightarrow + \gamma^0\chi_i$, then, because time reversal is anti-unitary, it acts on the Dirac fermion as $T:\psi \rightarrow \gamma^0 \psi^\dagger$. This reflects the fact that the symmetry group is $T\rtimes U(1)$.

A better way, as explained in \cite{sw}, is to consider CT. This forms the direct product $CT\times U(1)$, and acts on Dirac fermions in one of two ways,
\be {\rm CT}: \psi \rightarrow \pm \gamma^0 \psi\label{trdirac}\ee
In what follows, we will refer to CT simply as ``time reversal". The mod 16 anomaly is now given by
\be \nu_3 = 2\,(n_+ - n_-)\ \ {\rm mod\ 16}\label{nu3}\ee
where $n_\pm$ count the number of Dirac fermions that transform with a $\pm$ sign under time reversal \eqn{trdirac}. 

A quadratic mass term for fermions  -- whether Majorana $\psi_1\psi_2$ or Dirac $\bar{\psi}_1\psi_2$ -- breaks time reversal symmetry if both fermions transform with the same sign under CT.  In $d=2+1$ dimensions, the analog of symmetric mass generation is a mechanism which gaps 16 Majorara fermions,  or 8 Dirac fermions, all transforming in the same way under CT, while preserving time reversal\footnote{A simple theory with this property was described by Witten in \cite{witten}. It consists of a $U(2)$ gauge theory, with four Dirac fermions transforming in the ${\bf 2}$ and a complex scalar $\phi$ transforming with charge $-2$ under $U(1) \subset U(2)$. (The scalar lives in the determinant line bundle.) The scalar couples through the Yukawa term
\be {\cal L}_{\rm Yuk} = \phi \, (\psi_1 \psi_2 + \psi_3\psi_4)\nn\ee
where the fermion-bilinears are singlets under $SU(2) \subset U(2)$.  If $\phi$ gets a vev, then the fermions are gapped but time reversal is broken because $\phi$ is odd under CT. However, one can construct a new time-reversal symmetry $({\rm CT})' = K\cdot CT$ where $K$ is part of the broken gauge symmetry, acting as $-1$ on the scalar $\phi$ and as $+i$ on the fermions. Hence, the theory is gapped, while preserving a time reversal. However, because of the extra factor of $K$, time reversal acts on the gapped spectrum as $({\rm CT})^{\prime 2} = +1$ rather than $(-1)^F$.}.

\subsection{An S-Confining Theory in  $d=2+1$}

In this section, we see what becomes of the 4d supersymmetric $SU(2)$  s-confining theory when compactified on a circle. After a deformation, we will argue that the theory is trivially gapped, while preserving time reversal. As a check, we compute the index $\nu_3$ and find that it does indeed vanish, mod 16\footnote{The Smith isomorphism provides a map between the 4d anomaly \eqn{nu4} associated to \Sfour and the 3d anomaly \eqn{nu3} associated to time reversal. Physically, this arises by realising a 3d theory on a domain wall inside the 4d theory \cite{smithwall}. In contrast, here we will perform a dimensional reduction of the 4d theory (together with a particular mass deformation) and it is not obvious how to directly relate the 3d and 4d anomalies. Instead, we will simply compute them.}.

 The starting point is the 4d s-confining theory
\be d=3+1,\;  {\cal N}=1,\; SU(2) \ {\rm with}\ 6 \ {\rm fundamental\ chirals}\nn\ee 
As we described in Section \ref{4dsec}, this theory  is known to confine and flows, at the origin of moduli space to a theory of massless mesons and baryons, transforming in the ${\bf 15}$ of the $SU(6)$ flavour symmetry \cite{nati1,nati2}.

We now compactify  on ${\bf S}^1$. As explained in \cite{ahiss,shlomo}, this generates a monopole superpotential, 
\be {\cal W}_{KK} = \eta Y\label{wy}\ee
with $Y$ the monopole operator and $\eta$ a fixed parameter, related to the 4d strong coupling scale and the radius of the circle. 

The next step is to turn on equal and opposite, real masses for the $5^{\rm th}$ and $6^{\rm th}$ quarks. This breaks the flavour symmetry $SU(6) \rightarrow SU(4) \times U(1) \times U(1)$, under which the quarks decompose as 
\be {\bf 6} \rightarrow {\bf 4}_{-1,0} + {\bf 1}_{2,1} + {\bf 1}_{2,-1}\nn\ee
The singlets become heavy and decouple from the low-energy dynamics. The addition of real masses has a further, more subtle effect, shown in \cite{shlomo}: it kills the non-perturbative superpotential \eqn{wy}. The upshot is that we are left with an $SU(2)$ gauge theory coupled to 4 fundamental chiral multiplets

To understand the low-energy dynamics of this theory, we can follow the fate of the 4d low-energy meson fields when compactified on ${\bf S}^1$. 
Upon deforming by the real masses, the meson fields $M^{ij}$ with $i,j=1,\ldots,6$ decompose as
\be {\bf 15} \rightarrow {\bf 6}_{-2,0} + {\bf 4}_{1,1} + {\bf 4}_{-1,-1} + {\bf 1}_{4,0}\nn\ee
where the two fields in the ${\bf 4}$  become heavy and decouple. We are left with a free theory, consisting of a ${\bf 6}$ under the $SU(4)$ flavour symmetry and a singlet. The ${\bf 6}$ arises as composite mesons, $M^{ij} = Q^i Q^j$ with $i,j=1,\ldots,4$;  the singlet is dual to the monopole operator $M^{56}=Y$.

Finally, to gap these fields we play the same game that we saw in Section \ref{4dsec}; we return to the UV 3d gauge theory and add extra gauge singlets which we denote as $\Phi$ and $\Phi_0$. The end result is that we have an $SU(2)$ gauge theory with field content and symmetries given by
\begin{table}[H]
\begin{center}
\begin{tabular}{|c||c||c|c|c|}
  \hline 
   Field & $SU(2)_{\rm gauge}$ & $SU(4)$ & $U(1)_A$  & $U(1)_R$\\
 \hline
 $Q$  & ${\bf 2}$ & ${\bf 4}$  & $-1$  & $1/2$\\ 
  $\Phi$  & ${\bf 1}$ & ${\bf 6}$  & $2$  & $1$\\ 
$\Phi_0$ &  ${\bf 1}$ & ${\bf 1}$  & $-4$ & $2$\\ 
\hline
$Y$ & ${\bf 1}$ & ${\bf 1}$ & $+4$ & 0\\
\hline
 \end{tabular}
 \end{center}
\end{table}

\noindent
The quantum numbers of the monopole operator $Y$ are in accord with quantisation of the zero modes in the background of the monopole \cite{ahiss}. We write $Y$ below the line because, as a disorder operator, it should not be included in the accounting of the anomaly. That would be double-counting. We also  add a superpotential, consistent with all symmetries
\be {\cal W}_{UV} = \Phi_{ij} Q^i Q^j + \Phi_0 Y\label{w3d}\ee
Note that the superpotential includes the monopole operator. Importantly, and in contrast to the superpotential \eqn{wy},  $\Phi_0$ is dynamical. Its role is to remove the monopole operator from the chiral ring. 

From the discussion above, we this theory flows to a collection of free meson fields, coupled to the singlets $\Phi$ and $\Phi_0$ through the superpotential 
\be {\cal W}_{IR} = \Phi_{ij}M^{ij} + \Phi_0M^{56}\nn\ee
We see that the theory is gapped, with no topological sector, and time reversal in tact. 

\subsubsection*{The Mod 16 Anomaly}

Since this theory is gapped while preserving time reversal invariance, general considerations mean that its mod 16 anomaly must vanish. Indeed, as we now show, this is the case and can be viewed as symmetric mass generation for the 16 Majorana fermions that sit in $Q$. 

First, we pick a choice for transformation of the gaugino under time reversal, say
\be {\rm CT}: \lambda \rightarrow + \gamma^0 \lambda\nn\ee
Each chiral multiplet is either odd or even under CT. Expanding a generic chiral multiplet gives
\be \Phi = \phi + \theta \psi + \ldots\nn\ee
The superspace coordinate $\theta$ has the same transformation as the gaugino. This means that the transformation of the fermion $\psi$ is given by
\be {\rm CT}: \psi \rightarrow \pm \gamma^0 \psi\nn\ee
where the $+$ sign arises if the associated scalar $\phi$ is odd under $CT$, and the minus sign arises if the scalar $\phi$ is even. 

The superpotential must be odd if it is to preserve time reversal (because it must cancel the minus sign coming from the superspace measure $d^2\theta$). Indeed, this makes sense: we know that a mass ${\cal W} \sim \Phi^2$ breaks time reversal, while a Yukawa self-coupling ${\cal W} \sim \Phi^3$ preserves time reversal but only if the scalar $\phi$ is odd.

A glance at the superpotential \eqn{w3d} shows that it doesn't matter whether $Q$ are even or odd. This is because $Q\rightarrow -Q$ is part of the $SU(2)$ gauge symmetry.

Meanwhile, $\Phi$ must be odd. That leaves $\Phi_0$, whose behaviour under time-reversal is dictated by  $Y$. This, in turn, can be determined from the fermionic zero modes of the monopole operator. It picks up two  complex zero modes from the gaugino $\lambda$,  and four zero modes from the $Q$. These latter can be lifted by an obstruction bundle, and the instanton has the potential to contribute to  a $\langle \lambda\lambda \rangle$ correlation function. Since $\lambda$ is odd, $Y$ too must be odd. 

We can reach this same conclusion from the dual picture. Recall that, to derive the 3d theory from compactification, we added equal and opposite, real mass terms for the $5^{\rm th}$ and $6^{\rm th}$ quarks. Such real masses break time reversal. However, time reversal can be restored if accompanied by an exchange of the these two quarks, $Q^5\leftrightarrow Q^6$. This does not affect the mesons $M^{ij}$ with $i,j=1,\ldots 4$, and these remain even under time reversal. However, the final massless meson $M^{56}$ picks up a relative minus sign, and is odd under time reversal.

Both of the arguments above tell us that $M^{56} = Y$ is odd under time reversal. So $\Phi_0$ must be even.
We learn that the mod 16 anomaly of our system is 
\be \nu_3 = 2\,(3 + 6 -1) \pm 16 \nn\ee
where the $\pm$ sign depends on the choice of time reversal assigned to $Q$. The index with either choice of sign  must vanish mod 16. And, indeed, it does.

\appendix

\section{Appendix: No  Symmetric Mass Generation in Wess-Zumino Models}\label{indexsec}

In this paper, we have studied examples of symmetric mass generation induced by gauge dynamics. In $d=3+1$ dimensions, this is the only option available to drive strong coupling effects in the infra-red. However, in lower dimensions it may be possible to induce symmetric mass generation without gauge interactions. Indeed, the original work of Fidkowski and Kitaev \cite{fk} can be viewed as gapping 8 Majorana fermions while  preserving a ${\bf Z}_2$ symmetry that suffers a mod 8 anomaly \cite{ryu,qi}: it achieves this by invoking 4-fermion terms, without gauge interactions.

We could then ask: is it possible to find a supersymmetric counterpart to this interesting strongly coupled phenomenon. In this appendix, we show that the answer is no, at least within the context of theories of four supercharges, {\it i.e} ${\cal N}=2$ in 3d or ${\cal N}=(2,2)$ in 2d.

We consider a collection of  chiral superfields transforming in some representations of the global symmetry group $G$, together with a  superpotential consistent with such a symmetry. We assume that there exists a supersymmetric vacuum that leaves $G$ unbroken. Furthermore, we assume that $G$ is chiral, in the sense that it prohibits a supersymmetric mass term. In this case, no such term will be generated along the RG flow. The question  is whether strong supersymmetric dynamics can, nonetheless,  gap the model in the IR.

In principle, this is possible. For example, a gap may emerge if the IR theory has an effective description in terms of fields that are composite operators of the UV model. Then the symmetries might be consistent with mass terms for these fields which would appear as higher dimensional superpotentials in the UV.  Indeed, we witnessed this kind of behaviour in the gauge theories discussed in the bulk of the paper. Here we show that this is not possible in the absence of gauge interactions.

Our argument proceeds by use of the 4d supersymmetric index \cite{Kinney:2005ej}. Of course, in 4d any Wess-Zumino model is infra-red free. However, very similar index calculations also hold for the dimensional reduction to 3d and 2d where, a priori, one might have expected more interesting dynamics to occur. We consider $N$ chiral superfields $\Phi_i$ with R-charges $R_i$, and charges $Q^i_a$ under the Cartan subalgebra of $G$, where $i=1,\ldots,N$ labels the superfields and $a=1,\ldots,{\rm rank\,G}$ labels the Cartan element $U(1)\subset G$. We assume that none of the fields has a mass term of the form $\Phi^2$ as an index of such a field is trivially $1$.
The index is given by a product of elliptic Gamma functions, of the form
\be
{\cal I}=\prod_{i=1}^N\Gamma_e\left[(qp)^{R_i/2} \prod_{a=1}^{{\rm rank}\, G} y_a^{Q^i_a};q,p\right]\nn
\ee 
with $q$, $p$ and $y_a$ fugacities, and with  conventions that largely follow \cite{Dolan:2008qi,Rastelli:2016tbz}. The elliptic Gamma function can be expressed as 
\be
\Gamma_e(z;q,p) = \text{PE}\left[\frac{z- q\,p\, z^{-1}}{(1-q)(1-p)}\right]\nn\ee
where the plethystic exponential is defined by
\be  \text{PE}\left[f(x,y,\cdots)\right]=\exp\left[\sum_{l=1}^\infty \frac1l\,f(x^l,y^l,\cdots)\right]\nonumber
\ee 
If the theory is  gapped, it has
\be 
{\cal I}=1\nn
\ee 
This reflects the fact that, in the IR, we have only a single state which is the  supersymmetric vacuum. Clearly we must have a product of elliptic Gamma functions that equals one. One way in which this can happen is if the Gamma functions cancel in pairs, so that
\be
\Gamma_e \left[(qp)^{R_i/2} \prod_{a=1}^{{\rm rank}\, G} y_a^{Q^i_a};q,p\right] \Gamma_e \left[(qp)^{R_j/2} \prod_{b=1}^{{\rm rank}\, G} y_b^{Q^j_b};q,p\right] =1
\nn\ee 
This requires $i$ and $j$ to have the property that $R_i+R_j=2$ and $Q^i_a+Q^j_a=0$. In this case, the corresponding fields $i$ and $j$ can be paired together and lifted through a mass term. However, the assumption that $G$ is chiral means that no such mass terms are possible. This means that we should search for a more creative way for the Gamma functions to cancel. We now show that no such creative way exists.

To see this, note that if the product of  elliptic Gamma functions is equal to one then 
setting the fugacities for all the $U(1)$ symmetries to $y_a=1$  we must have (defining $x=(q p)^{1/2}$)

\be
\text{PE}\left[\frac1{(1-q)(1-p)}\, \sum_{i=1}^N\left(x^{R_i}-x^{2-R_i}\right)\right]=1\,,\nonumber
\ee
which implies 
\be
\sum_{i=1}^N x^{R_i-1} =\sum_{i=1}^N x^{1-R_i}
\nn\ee 
This should hold for arbitrary value of $x$. It can happen only if there is a permutation $\sigma$ of $\{1,...,N\}$  such that $R_i=2-R_{\sigma(i)}$.  A similar argument in the presence of fugacities $y_a$ for the $U(1)$ symmetries ensures that $Q_a^i+Q_a^{\sigma(i)}=0$, confirming that the only possible solution is that in which the chiral multiplets cancel in pairs.

Although we have phrased the discussion above in terms of the 4d index, the argument can be extend to the supersymmetric indices of ${\cal N}=2$ theories in $3d$ \cite{Kim:2009wb,Willett:2016adv} and ${\cal N}=(2,2)$ theories in $2d$ \cite{Benini:2013nda,Benini:2013xpa}.

The story above assumed no gauging of a symmetry, neither continuous nor discrete. Introducing such gauging provides a loophole to the argument above, because the index now involves discrete sums or continuous integrals of some special functions\footnote{For example in 2d with discrete gauging see \cite{Aharony:2017adm,Hori:2011pd}
.}. Though a product of these special functions can be equal to one only if there is a mass term, sums (or integrals) of products can be equal to one even without mass terms, as is the case for the examples of symmetric mass generation given in the bulk of the paper. 

Nonetheless, the end result is perhaps a little surprising. The Fidkowski-Kitaev mechanism of symmetric mass generation  in d=1+1 does not rely on gauge interactions. It would appear that this is an example of a strongly coupled phenomenon that does not have a counterpart in the supersymmetric world.


\acknowledgments
We are grateful to  Pietro Benetti Genolini, Philip Boyle Smith, Joe Davighi, Avner Karasik, Nakarin Lohitsiri, Carl Turner for many conversations on symmetric mass generation, and the numbers 8 and 16.
DT is supported by the STFC consolidated grant ST/P000681/1, a Wolfson Royal Society Research Merit Award holder and a Simons Investigator Award. The research of SSR is supported in part by Israel Science Foundation under grant no. 2289/18, by I-CORE  Program of the Planning and Budgeting Committee, by a Grant No. I-1515-303./2019 from the GIF, the German-Israeli Foundation for Scientific Research and Development, and by BSF grant no. 2018204.

\end{document}